\documentclass[prd,byrevtex
,preprint
,showkeys
,showpacs]{revtex4}
\usepackage{amsmath,amsfonts}
\usepackage[scaled]{helvet}
\usepackage{bm}
\usepackage[dvipdfm]{graphicx}
\usepackage{float}
\newcommand{\bs}[1]{\boldsymbol{#1}}

\newcommand{\al}{\alpha}
\newcommand{\del}{\delta}
\newcommand{\ep}{\epsilon}
\begin{document}
\title{Quantum Energy Teleportation with a Linear Harmonic Chain}
\author{Yasusada Nambu}
\email{nambu@gravity.phys.nagoya-u.ac.jp}
\affiliation{Department of Physics, Graduate School of Science, Nagoya 
University, Chikusa, Nagoya 464-8602, Japan}
\author{Masahiro Hotta} 
\email{hotta@tuhep.phys.tohoku.ac.jp}
\affiliation{Department of Physics, Faculty of Science, Tohoku
  University, Sendai 980-8578, Japan}
\date{October 14, 2010}
\begin{abstract}
A protocol of quantum energy teleportation is proposed for a
one-dimensional harmonic chain. A coherent-state positive
operator-valued measure (POVM) measurement is
performed on coupled oscillators of the chain in the ground state
accompanied by energy infusion to the system. This measurement
consumes a part of the ground state entanglement. Depending on the
measurement result, a displacement operation is performed on a distant
oscillator accompanied by energy extraction from the zero-point
fluctuation of the oscillator. We find that the amount of consumed
entanglement is bounded from below by a positive value that is
proportional to the amount of  teleported energy.
\end{abstract}
\maketitle

\section{Introduction}

Recently, it has been reported that energy can be transported by local
operations and classical communication (LOCC) while retaining local
energy conservation and without breaking causality ( for spin
systems~\cite{HottaM:PLA372:2008,HottaM:JPSJ78:2009,HottaM:PLA374:2010},
for trapped ion systems~\cite{HottaM:PRA80:2009}, and for quantum field
systems~\cite{HottaM:PRD78:2008,HottaM:JPAMT43:2010,HottaM:PRD81:2010}).
Such protocols are called quantum energy teleportation (QET) and are
based on ground-state entanglement of many-body quantum systems. By
performing a local measurement on a subsystem $A$ of a many-body system
in the ground state, information about the quantum fluctuation of $A$
can be extracted. During this measurement, some amount of energy is
infused into $A$ as QET energy input, and the ground state entanglement
gets partially broken. The measurement result is announced to another
subsystem $B$ of the many-body system by the classical protocol. Using
this information, energy can be extracted from $B$ by performing a local
operation on $B$ dependent on the announced measurement data. The root
of the protocols is a correlation between the measurement information
of $A$ and the quantum fluctuation of $B$ via the ground state
entanglement. The information about $A$ enables us to partially know what
kind of zero-point fluctuation of $B$ is realized. Thus, using this
information, we can select a good operation on $B$ for the energy
extraction.

In general, we are able to make a better strategy for a task by
obtaining more information. Hence, for the QET case, it sounds
plausible to imagine that more information about the quantum
fluctuation of $B$ is obtained by measurements of $A$, more energy can be
teleported from $A$ to $B$. If we consume a large amount of ground-state
entanglement between fluctuation of $A$ and fluctuation of $B$ during the
measurement of $A$, it is naturally expected that much information about
the post-measurement state of $B$ is included in $A$'s measurement
result. Therefore, there should exist some qualitative relationship
between the breaking of entanglement by measurement and the amount of
 teleported energy of the optimal operation of $B$.

While the basic protocol of QET has been investigated for several model
systems, the analysis of QET for many-body systems has not been
development yet. This is because these quantities are difficult to obtain
analytically for many-body systems. So far, the relation between
teleported energy and entanglement breaking has been investigated only for
the minimal model of QET~\cite{HottaM:PLA374:2010} which consists of two
qubits.  In this paper, we aim to establish this relation of QET using
a one-dimensional harmonic chain.  There are many works investigating
the feature of entanglement in harmonic chain
models~\cite{AudenaertK:PRA66:2002,BoteroA:PRA70:2004,KoflerJ:PRA73:2006,NambuY:PRD78:2008,MarcovitchS:PRA80:2009}. The
conventional method to analyze the entanglement in a harmonic chain is
to introduce two spatial regions containing sites of the harmonic
chain and consider the bipartite entanglement between these
regions. Previous analysis shows that the zero temperature ground
state (vacuum state) of the harmonic chain is entangled. Thus this
entanglement can become a resource for QET.  As we are not interested
in the dynamical aspect of QET in the present analysis, we assumed
that the speed of classical communication is infinity and our protocol
can be treated as the non-relativistic one.

For a harmonic chain, we first consider the setting that groups $A$ and
the $B$ consist of a single site. We investigate the relation between the
amount of teleported energy and the quantum mutual information by changing
the distance between $A$ and $B$. Then we consider the setting with a
block of coupled harmonic oscillators as $A$ and its complementary set
as $B$.  Increasing the number of measured oscillators in $A$, we
calculate both the entanglement breaking between $A$ and $B$ and the
amount of energy teleported from $A$ to $B$ via a QET protocol. As a
measure of entanglement, we use the logarithmic
negativity~\cite{VidalG:PRA65:2002}. An explicit inequality is given
such that the breaking of entanglement is lower bounded by a positive
value that is proportional to the amount of teleported energy. This
shows how much entanglement is required to teleport energy in the
harmonic chain.

The paper is organized as follows. In Sec.~II, we introduce the
harmonic chain model and we prepare a formula for the positive
oeprator-valued measure (POVM) measurement for the harmonic chain in
Sec.~III. We present our numerical result in Sec.~IV and Sec.~V is
devoted to a summary.

\section{Harmonic chain model}
We consider a one-dimensional harmonic chain with $N$ sites. We
assume that $N$ is an even number.  The Hamiltonian is
\begin{equation}
  H=\frac{1}{2}\sum_{j=1}^N\left(p_j^2+q_j^2-\al
    q_jq_{j-1}\right),\qquad (j=1,\cdots,N)
\end{equation}
where we assume a periodic boundary condition for $q_j,p_j$ and a
positive parameter $\al\neq 1$ is introduced to regularize the
infrared divergence which appears in the correlation matrix.  $\al=1$
corresponds to the critical case (massless limit) of the harmonic
chain. This criticality allows us to teleport energy to a distant
point, decreasing the amount of energy obeying a power law decay with
respect to the distance. As we will show, QET is possible for
the non-critical case. However, the amount of teleported energy decreases
exponentially with increasing distance and QET is not so
effective. Thus, it is important to investigate QET with near-critical
harmonic chains. The quantized canonical variables are
\begin{align}
  & q_j=\frac{1}{\sqrt{N}}\sum_{k=0}^{N-1}\left(f_k\hat a_k+f_k^*\hat
    a_{N-k}^\dag\right)e^{i\theta_k j}, \\
  & p_j=\frac{1}{\sqrt{N}}\sum_{k=0}^{N-1}(-i)\left(g_k\hat
    a_k-g_k^*\hat a_{N-k}^\dag\right)e^{i\theta_k j},\quad
  \theta_k=\frac{2\pi k}{N} \notag
\end{align}
where $f_k,g_k$ are mode functions
\begin{equation}
 f_k=\frac{1}{\sqrt{2\omega_k}}\,e^{-i\omega_k t},\quad
 g_k=\sqrt{\frac{\omega_k}{2}}\,e^{-i\omega_k t},\quad \omega_k^2=1-\al\cos\theta_k
\end{equation}
and the creation and annihilation operators satisfy
$$
 \left[\hat a_k,\hat a^\dag_{k'}\right]=\del_{kk'}.
$$
The two point correlation functions with respect the ground  state
satisfying $\hat a_k|g\rangle=0$ are
\begin{align}
  &\langle
  q_iq_j\rangle=\frac{1}{N}\sum_{k=0}^{N-1}\frac{1}{2\omega_k} 
  \cos\left[(i-j)\theta_k\right]=
  g_{|i-j|}=G_{ij}, \label{eq:G}\\
  &\langle
  p_ip_j\rangle=\frac{1}{N}\sum_{k=0}^{N-1}\frac{\omega_k}{2}\cos\left[(i-j)\theta_k\right]
  = h_{|i-j|}=H_{ij}, \label{eq:H}\\
  &\langle
  q_ip_j\rangle=\frac{i}{2N}\sum_{k=0}^{N-1}e^{i\theta_k(i-j)}=\frac{i}{2}\del_{ij},\\
  &\langle q_i\rangle=\langle p_i\rangle=0, \notag
\end{align}
where we have introduced the matrices $G$ and $H$ satisfying the relation
\begin{equation}
  (GH)_{ij}=\frac{1}{4}\,\del_{ij}.
\end{equation}
As the ground state is a Gaussian state, it can be completely determined
using these two point correlation functions.
\section{POVM measurement and  energy teleportation}
We briefly review the protocol of QET (Fig.~\ref{fig:protocol}).  In
these figures, quantum fluctuations in the harmonic chain are
schematically shown as broken lines. The dashed horizontal lines
represent the magnitude of quantum fluctuation of the state. The protocol
consists of the following steps:
\begin{figure}[H]
  \centering
  \includegraphics[width=0.7\linewidth,clip]{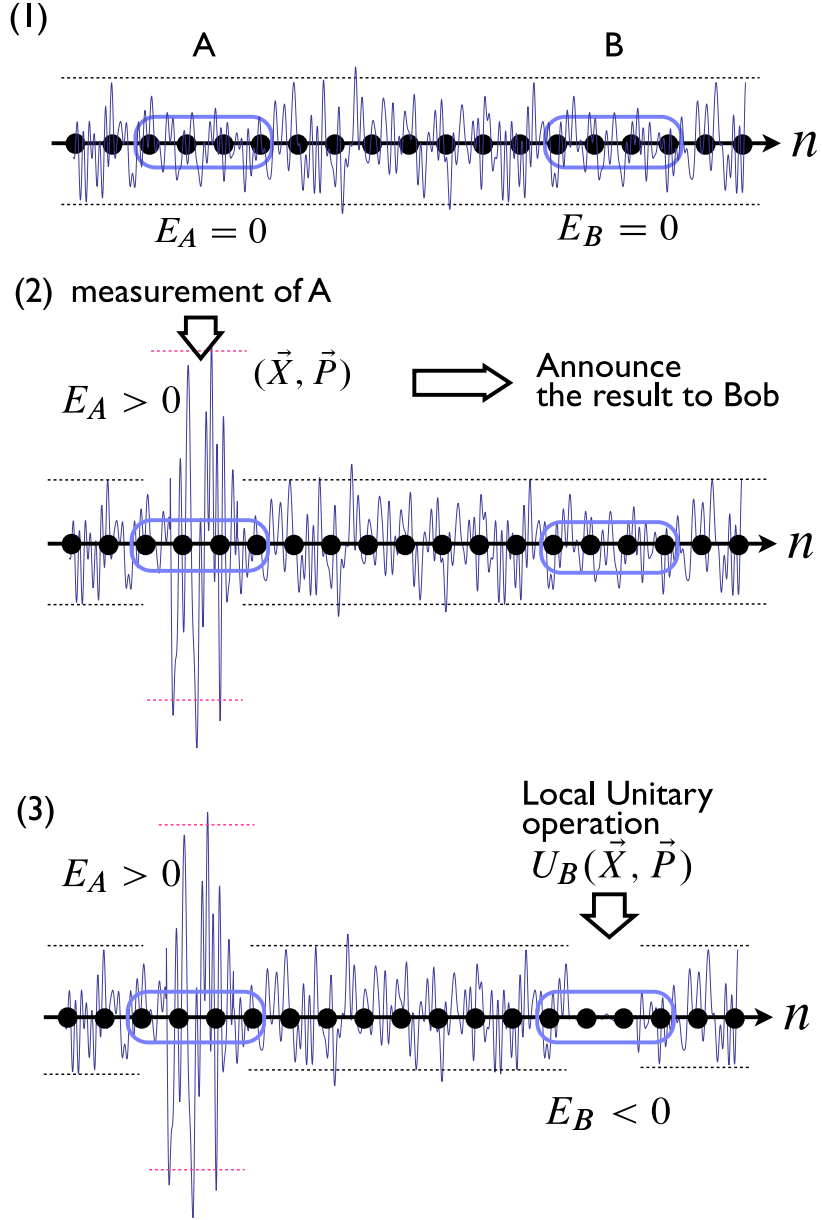}
  \caption{The protocol of QET with a harmonic chain. (1) We prepare
    the ground state of a harmonic chain and introduce two parties $A$
    and $B$. The local energy of $A$ and $B$ is zero. (2) The measurement of
    $A$ increases the energy of the group $A$. The result of the
    measurement is announced to Bob via classical communication. (3)
    Depending on the announced measurement result, Bob performs a
    local unitary operation on the group $B$, which reduces the energy
    of the group $B$ so that it becomes negative. The dashed horizontal lines
    represent the magnitude of quantum fluctuation of the state.}
  \label{fig:protocol}
\end{figure}
\noindent
(1) We prepare the ground state of the harmonic chain and define
groups $A$ and $B$ of coupled oscillators in the system. $A$ and $B$ consist
of some number of sites.  The groups $A$ and $B$ are entangled due to the
entanglement of the ground state quantum fluctuation of the harmonic
chain.  This entanglement enables us to extract a positive energy via
a protocol of QET.

(2) Alice makes a local measurement of quantum fluctuations of sites
in the group $A$. Because the post-measurement state is not the ground
state, it is an excited state with positive energy. As some
amount of energy is infused through this measurement procedure, the
energy of the group $A$ after the measurement becomes positive provided
that the ground state energy before the measurement is chosen to be
zero. By this measurement, Alice obtains measurement results for
sites in the group $A$. She announces these values to Bob via classical
communication.

(3) Depending on the announced measurement results, Bob performs a
local unitary operation on sites in the group $B$. By choosing the
operation suitably to suppress the zero-point fluctuation of $B$, it can
be shown that the local energy of the group $B$ takes a negative value lower
than the ground-state one. Thus we can extract positive energy from
the system.

In our numerical investigation of QET for the harmonic chain, we apply
a POVM measurement~\cite{NielsenMA:CUP:2000} to sites in the group $A$
(see Fig.~\ref{fig:setting1} and \ref{fig:setting2}).  The
measurement reduces the entanglement between the two groups. What we
want to know is the relation between the amount of  entanglement
breaking due to the measurement and the extracted energy via the
protocol of QET.

\subsection{The state before and after the measurement}
We assume that the state of the system before the measurement
is the ground state. In the coordinate representation, the state is
given by
\begin{equation}
  \rho_0=\langle\vec q|g\rangle\langle g|\vec q'\rangle,\quad\vec q=(q_1,\cdots,q_N)^T
\end{equation}
where
$$
\langle
\vec q|g\rangle\propto\exp\left[-\frac{1}{4}\sum_{i,j=1}^Nq_i(G^{-1})_{ij}q_j\right]
=\exp\left[-\sum_{i,j=1}^{N}q_iH_{ij}q_j\right].
$$
We apply a coherent-state POVM measurement to $2\ell+1$ sites in the
group $A$.  To define the POVM adopted in our analysis, let us introduce
the following annihilation operator for the canonical variable
$(q_n,p_n)$ of the site $n$ in the group $A$:
$$
 \hat b_n=\sqrt{\frac{\omega}{2}}\,\hat
 q_n+\frac{i}{\sqrt{2\omega}}\,\hat p_n,\quad 1\le n\le 2\ell+1.
$$
We define the ground state $|0_n\rangle$ with respect to the operator
$\hat b_n$ as 
$$
 \hat b{}_n|0_n\rangle=0
$$
and the coherent state defined by $\hat b_n^\dag$ as
\begin{equation}
 \label{eq:coherent}
 |X_n, P_n\rangle=e^{-|c_n|^2/2}\exp\left(c_n\,{\hat
   b}{}_n^\dag\right)|0_n\rangle,\quad
 c_n=\sqrt{\frac{\omega}{2}}\,X_n+\frac{i}{\sqrt{2\omega}}\,P_n,
\end{equation}
where $X_n$ and $P_n$ are real numbers and  are classical
amplitude of the coherent state.  The expectation values of site
variables with respect to this coherent state are
\begin{align}
 &\langle X_n,P_n| q_n|X_n,P_n\rangle=X_n,\quad  \langle X_n,P_n|
 p_n|X_n,P_n\rangle=P_n,\\
 &\langle X_n,P_n| q_n^2|X_n,P_n\rangle=X_n^2+\frac{1}{2\omega},\quad 
\langle X_n,P_n| p_n^2|X_n,P_n\rangle=P_n^2+\frac{\omega}{2}.\notag
\end{align}
In the coordinate representation, the coherent state can be written
\begin{equation}
  \langle q_n|X_n,P_n\rangle
  =\left(\frac{\omega}{\pi}\right)^{1/4}\exp\left[-\frac{\omega}{2}(q_n-X_n)^2
+i P_n(q_n-X_n)\right].
\end{equation}
The measurement operator for the POVM  is given by
\begin{equation}
  M_\ell(\vec X,\vec P)=\frac{1}{(\sqrt{2\pi})^{2\ell+1}}
\prod_{n=1}^{2\ell+1}|X_n,P_n\rangle\langle
  X_n,P_n|,\quad 1\le \ell\le\frac{N}{2}-2
\end{equation}
where
$$
 \vec X=(X_1,\cdots,X_{2\ell+1})^T,\quad \vec P=(P_1,\cdots,P_{2\ell+1})^T.
$$
The POVM is defined by
\begin{equation}
 \Pi_\ell(\vec X,\vec P)=M_\ell^{\dag}(\vec X,\vec P) M_\ell(\vec X,\vec P)=
 \frac{1}{(2\pi)^{2\ell+1}}\prod_{n=1}^{2\ell+1}|X_n,P_n\rangle\langle X_n,P_n|
\end{equation}
and satisfies the following completeness relation for the measurement operator:
$$
 \int\left(\prod_{n=1}^{2\ell+1}dX_ndP_n\right)\Pi_\ell(\vec X,\vec P)=I.
$$
If we obtain  the value $(\vec X,\vec P)$ as the result of the POVM measurement, the
state after the measurement can be written as
\begin{equation}
\label{eq:rho-after}
\rho_\ell(\vec X,\vec P)=\prod_{n=1}^{2\ell+1}\langle
X_n,P_n|g\rangle\langle g|X_n,P_n\rangle
\otimes\prod_{n'=1}^{2\ell+1}|X_{n'},P_{n'}\rangle\langle X_{n'},P_{n'}|.
\end{equation}
After the measurement, the states for the measured sites become the
coherent states \eqref{eq:coherent} and states for other sites are
computed by acting the measurement projection operator on the ground
state.  To obtain the explicit form of the state after the POVM
measurement, we introduce the following vector notation for the site
variables:
$$
(q_n)=\begin{pmatrix} \vec q_\ell^M \\ \vec q_\ell\end{pmatrix},\quad
(p_n)=\begin{pmatrix} \vec p_\ell^M \\ \vec p_\ell\end{pmatrix}.
$$
The $(2\ell+1)$-dimensional vectors $\vec q_\ell^M, \vec p_\ell^M$
represent variables for the sites whose values are measured. Then,
with respect to these bases, the correlation matrices \eqref{eq:G} and
\eqref{eq:H} for site variables can be written
$$
H=\begin{pmatrix} L_\ell & K_\ell \\ K_\ell^T & H_\ell \end{pmatrix},\quad
G=\begin{pmatrix} C_\ell & D_\ell \\ D_\ell^T & G_\ell \end{pmatrix},
$$
where $L_\ell, C_{\ell}$ denote the $(2\ell+1)\times(2\ell+1)$
correlation matrices for the measured sites. Using this notation, the
ground state projected to the coherent state \eqref{eq:coherent} is given by
\begin{align}
  \langle\vec X,\vec P|g\rangle &\propto\exp\Biggl[
  -\vec
  q_\ell^T\left(H_\ell-K_\ell^T\left(L_\ell+\frac{\omega}{2}\right)^{-1}K_\ell\right)
  \vec q_\ell
  +\vec q_\ell^TK_\ell^T\left(L_\ell+\frac{\omega}{2}\right)^{-1}(i\vec
  P-\omega\vec X) \notag \\
  &\qquad\qquad -\frac{\omega}{2}\vec
  X^T\left(1-\frac{\omega}{2}\left(L_\ell+\frac{\omega}{2}\right)^{-1}\right)\vec
  X-\frac{1}{4}\vec P^T\left(L_\ell+\frac{\omega}{2}\right)^{-1}\vec
  P\\
  &\qquad\qquad
  +\frac{i}{2}\vec
  P^T\left(1-\frac{\omega}{2}\left(L_\ell+\frac{\omega}{2}\right)^{-1}\right)\vec
  X
  +\frac{i}{2}\vec
  X^T\left(1-\frac{\omega}{2}\left(L_\ell+\frac{\omega}{2}\right)^{-1}\right)\vec
  P\Biggr]. \notag
\end{align}
The two-point correlation functions for fluctuations $\Delta q_i=q_i-\langle
q_i\rangle, \Delta p_i=p_i-\langle p_i\rangle $ after the measurement
are obtained by taking the expectation value with respect to the state
\eqref{eq:rho-after}. The non-zero correlation functions are
\begin{alignat}{2}
 &i,j\in\text{(measured site in $A$)}:&\quad 
&\langle \Delta q_i\Delta q_j\rangle=\frac{\del_{ij}}{2\omega},\quad\langle
 \Delta p_i\Delta p_j\rangle=\frac{\omega}{2}\del_{ij},\notag \\
 &&&\langle \Delta p_i\Delta q_i+\Delta q_i\Delta p_i\rangle=0 \notag \\
 &i,j\notin \text{(measured site in $A$)}:&\quad &\langle\Delta q_i\Delta
 q_j\rangle=\frac{1}{4}\left(M^{-1}\right)_{i-(2\ell+1),j-(2\ell+1)},
  \\
 &&&\langle \Delta p_i\Delta p_j\rangle=\left(M\right)_{i-(2\ell+1),j-(2\ell+1)} \notag
\end{alignat}
where the matrix $M$ is defined by
\begin{equation}
  M=H_\ell-K_\ell^T\left(L_\ell+\frac{\omega}{2}\right)^{-1}K_\ell.
\end{equation}
The other two-point functions are zero.  These covariances are independent
of the result of the measurement $(\vec X, \vec P)$.  The probability
distribution to obtain the measured value $(\vec X, \vec P)$ is given by
\begin{align}
  \langle g|\Pi_{\ell}(\vec X,\vec P)|g\rangle&=\int d\vec
  q_\ell|\langle g|\vec X,\vec P\rangle|^2  \notag \\
&\propto
  \exp\left[-\frac{1}{2}\vec
    P^T\left(L_\ell+\frac{\omega}{2}\right)^{-1}\vec P-\frac{1}{2}\vec
    X^T\left(C_\ell+\frac{1}{2\omega}\right)^{-1}\vec X\right].
\end{align}

\subsection{The optimized energy}
We define the energy of the lattice site $n$ as
\begin{equation}
  H_n=\frac{1}{2}\left[p_n^2+q_n^2-\frac{\al}{2}q_n(q_{n-1}+q_{n+1})-\ep\right]
\end{equation}
where the constant $\ep$ is chosen  to satisfy
\begin{equation}
 \langle g|H_n|g\rangle =0.
\end{equation}
After the measurement, we obtain the values $(\vec X,\vec P)$ of the
measured sites. Depending on  these measured values, we apply the following
local displacement operation to sites in the group $B$ which is composed of a
single site in our numerical setting ($n_B=N/2+\ell+1$)
\begin{equation}
  U_B(\vec X,\vec P)=\exp\left[i\left((\vec\theta\cdot\vec
      P)q_{n_B}-(\vec\phi\cdot\vec X)p_{n_B}\right)\right]
\end{equation}
where $\vec\theta$ and $\vec\phi$ are constant parameters to be
determined later by optimization. Then, the transformed state is
\begin{equation}
  \rho_\ell'=\int\prod_{n=1}^{2\ell+1}(dX_ndP_n)\,U_B(\vec X,\vec P)M_d(\vec X,\vec
  P)|g\rangle\langle g|M_d^\dag(\vec X,\vec P)U_B^\dag(\vec X,\vec P).
\end{equation}
The energy of the site $B$ with respect to this state is
\begin{align}
  \langle H_B\rangle&=\sum_{j=1}^{2\ell+1}\theta_j\langle
  g|p_jp_{n_B}|g\rangle+
  \sum_{j=1}^{2\ell+1}\phi_j\langle
  g|q_j\left(q_{n_B}-\frac{\al}{2}(q_{n_B+1}+q_{n_B-1})\right)|g\rangle
  \notag \\
  &\qquad
  +\frac{1}{2}\sum_{j,k=1}^{2\ell+1}\theta_j\left(\langle
    g|p_jp_k|g\rangle+\frac{\omega}{2}\del_{jk}\right)\theta_k
  +\frac{1}{2}\sum_{j,k=1}^{2\ell+1}\phi_j\left(\langle
    g|q_jq_k|g\rangle+\frac{1}{2\omega}\right)\phi_k \notag \\
  &=\frac{1}{2}\vec\theta^TT_p\,\vec\theta+\vec J_p\cdot\vec\theta
  +\frac{1}{2}\vec\phi^TT_q\,\vec\phi+\vec J_q\cdot\vec\phi
\end{align}
where we have introduced
\begin{align}
  &\left(T_p\right)_{jk}=h_{|j-k|}+\frac{\omega}{2}\,\del_{jk},\quad
   \left(T_q\right)_{jk}=g_{|j-k|}+\frac{1}{2\omega}\,\del_{jk}, \\
  &\left(\vec J_p\right){}_j=h_{|j-n_B|},\quad \left(\vec J_q\right){}_j=
  g_{|j-n_B|}-\frac{\al}{2}\left(g_{|j-(n_B-1)|}+g_{|j-(n_B+1)|}\right). \notag
\end{align}
By choosing the following values of the parameters
\begin{equation}
  \vec\theta=-T_p^{-1}\vec J_p,\quad \vec\phi=-T_q^{-1}\vec J_q,
\end{equation}
we obtain the following minimum (optimized) value of the site $B$'s energy
\begin{equation}
  \langle H_B\rangle_{\text{opt}}=-\frac{1}{2}\vec J_p^TT_p^{-1}\vec
  J_p-\frac{1}{2}\vec J_q^TT_q^{-1}\vec J_q<0.
  \label{eq:EB}
\end{equation}
The negative sign of $\langle H_B\rangle$ means that a positive amount
of energy is transferred from the chain to outside, obeying local
energy conservation.

\subsection{Bipartite entanglement and mutual information}
To investigate the relation between the amount of extractable
energy via QET and the entanglement breaking due to the measurement,
we consider the bipartite entanglement between the group $A$ and the
group $B$. For this purpose, we use the logarithmic negativity $E_N$ as
our measure of entanglement. The logarithmic negativity provides an
upper bound on the efficiency of distillation of entanglement~\cite{VidalG:PRA65:2002}.

The canonical variables of our bipartite system are
$$
 \vec\xi^T=\begin{pmatrix}\vec q_A,  & \vec p_A, & q_B, &
   p_B\end{pmatrix},\quad
 \left[\xi_j,\xi_k\right]=i\Omega_{jk},\quad j,k=1,\cdots,2N
$$
where $\bs{\Omega}$ is a $2N\times 2N$ matrix
$$
 \bs{\Omega}=\begin{pmatrix}\bs{J} & & \\
     &\ddots &\\
     & & \bs{J}\end{pmatrix},\quad\bs{J}=\begin{pmatrix}0 & 1 \\ -1 & 0\end{pmatrix}.
$$
As we are assuming a Gaussian state, the state is completely specified by the
following covariance matrix
\begin{equation}
  V_{jk}=\frac{1}{2}\langle\xi_j\xi_k+\xi_k\xi_j\rangle
-\langle\xi_j\rangle\langle\xi_k\rangle.
\end{equation}
The symplectic eigenvalue $\nu_n$ of the covariance matrix $\bs{V}$
satisfies the inequality
\begin{equation}
  \nu_n\ge\frac{1}{2}
\end{equation}
which is the generalization of the uncertainty relation.  The
logarithmic negativity is defined by the partially transposed
covariance matrix $\tilde{\bs{V}}$ obtained by reversing the sign of
party $B$'s momentum
\begin{equation}
  E_N\equiv-\sum_{n=1}^{N}\mathrm{min}\left[0,\log_2(2\tilde\nu_n)\right]
\end{equation}
where $\tilde\nu_n$ is the symplectic eigenvalue of the partially
transposed covariance matrix $\tilde{\bs{V}}$. If this quantity is
positive, the bipartite system is entangled, and we can use the
logarithmic negativity as the measure of the bipartite entanglement
between groups $A$ and $B$. The logarithmic negativity provides the
sufficient condition for entanglement between $A$ and $B$.

We also consider the quantum mutual information of $A$ and $B$ as a measure of
correlation between $A$ and $B$. For the Gaussian system with the
covariance matrix $\bs{V}$, in terms of symplectic eigenvalues
$\nu_n$, the von Neuman entropy of the system is given by~\cite{HolevoAS:PRA59:1999}
\begin{equation}
  S=\sum_nf(\nu_n),\quad
  f(x)=\left(x+\frac{1}{2}\right)\ln\left(x+\frac{1}{2}\right)
-\left(x-\frac{1}{2}\right)\ln\left(x-\frac{1}{2}\right).
\end{equation}
For the bipartite system with the group $A$ and $B$, the mutual information is defined by
\begin{equation}
  \label{eq:mutu}
  S_M=S(A)+S(B)-S(A+B).
\end{equation}
This quantity represents the total correlations including the quantum
and the classical correlation between the group $A$ and
$B$~\cite{HendersonL:JPAMG34:68996905}. A large value of $S_M$ implies
that $A$ and $B$ are strongly correlated and we are able to obtain much
information about $B$ just by measuring $A$. Thus, this is expected to assure
a large amount of teleported energy.

We numerically calculate the logarithmic negativity and the mutual
information.  The covariance matrix of the total system before the measurement is
\begin{equation}
  V_0=
  \begin{pmatrix}U_0^{(1,1)} & \hdots & U_0^{(1,N)}\\
    \vdots & \ddots & \vdots \\
    U_0^{(N,1)} & \hdots & U_0^{(N,N)}
  \end{pmatrix},\quad
  U_0^{(i,j)}=\begin{pmatrix}\langle\Delta q_i\Delta q_j\rangle & 0 \\
    0 & \langle \Delta p_i\Delta p_j\rangle 
    \end{pmatrix}=\begin{pmatrix} g_{|i-j|} & 0 \\ 0 & h_{|i-j|} \end{pmatrix}.
\end{equation}
The covariance matrix after the measurement is
\begin{equation}
  V^M=\begin{pmatrix} V_A^M & 0 \\ 0 & V_{AB} \end{pmatrix},
\end{equation}
where $V_A^M$ represents the covariance matrix for the measured sites
and is given by the $2(2\ell+1)\times 2(2\ell+1)$ diagonal matrix
\begin{equation}
  V_A^M=\begin{pmatrix} U_1 & & \\ & \ddots & \\ & &
    U_1\end{pmatrix},\quad
  U_1=\begin{pmatrix}\dfrac{1}{2\omega} & 0 \\ 0 & \dfrac{\omega}{2}\end{pmatrix};
\end{equation}
and $V_{AB}$ represents the covariance matrix for the un-measured sites,
\begin{equation}
  V_{AB}=\begin{pmatrix} U^{(1,1)} & \hdots & U^{(1,N-2d-1)} \\
    \vdots & \ddots & \vdots \\
    U^{(N-2d-1,1)} & \hdots & U^{(N-2d-1,N-2d-1)} \end{pmatrix},\quad
  U^{(i,j)}=\begin{pmatrix}
    \dfrac{1}{4}\left(M^{-1}\right)_{i,j} & 0 \\ 0 &
    \left(M\right)_{i,j}
    \end{pmatrix}.
\end{equation}
By use of  these expressions for the covariance matrices, the logarithmic
negativity and the mutual information before and after the measurement
can be obtained numerically.

\section{Numerical Result}
We numerically calculated the decrease of the logarithmic negativity
due to the measurement 
$$
 \Delta E_N=E_N(\text{before measurement})-E_N(\text{after measurement})
$$
and the optimized energy of the site $B$ obtained via the protocol of
QET. We consider the two different settings of numerical calculations
to establish the relation between the entanglement breaking and the
amount of teleported energy. As the parameter $\al$ of the harmonic
chain, we used the following four values:
\begin{equation}
  \al_1=0.90,\quad \al_2=0.95,\quad \al_3=0.99,\quad 
  \al_4=1-10^{-7}~\text{(critical)}.
\end{equation}
We have also investigated harmonic chains with smaller values of
$\al$, but the amount of teleported energy is reduced and QET is not
effective for small values of $\al$. We adopt these four values of
$\al$ as typical cases of our numerical calculation. The values
$\al_1, \al_2, \al_3$ represent  typical examples of the non-critical
case and $\al_4$ represents the critical case with a small cutoff parameter
$10^{-7}$ which is necessary to evaluate correlation matrices
numerically. As we will see later, the numerically obtained quantities
converge to the value with $\al_4$ as the value of $\al$ approaches 1.

\subsection{Setting 1}
\begin{figure}[H]
  \centering
  \includegraphics[width=0.4\linewidth,clip]{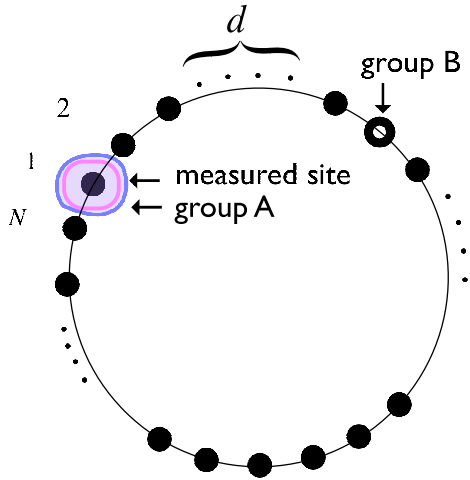}
  \caption{The setting 1 of our numerical calculation. We change the separation
    $d$ between sites $A$ and $B$.}
  \label{fig:setting1}
\end{figure}
\begin{figure}[H]
  \centering
  \includegraphics[width=0.5\linewidth,clip]{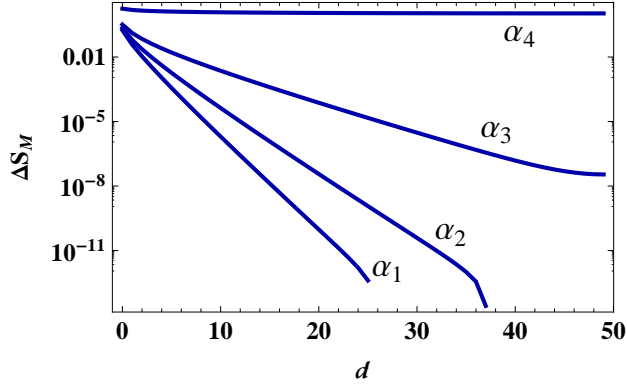}
  \caption{$d$ dependence of the decrease of the mutual
    information due to the measurement.  For the critical case
    $\al=\al_4$, this quantity behaves as $\Delta S_M\sim
    1.55d^{-0.11}$ for large $d$.}
  \label{fig:info-set1}
\end{figure}

As the first setting of our numerical calculation,  group $A$ and 
group $B$ consist of a single site (Fig.~\ref{fig:setting1}). We fix the
total site number $N=100$. We apply the POVM measurement to the site
$A$. By changing the separation $d$ between $A$ and $B$, we observe how the
entanglement between the two groups and the optimized energy
\eqref{eq:EB} of the site $B$ change.
\begin{figure}[H]
  \centering
  \includegraphics[width=0.45\linewidth,clip]{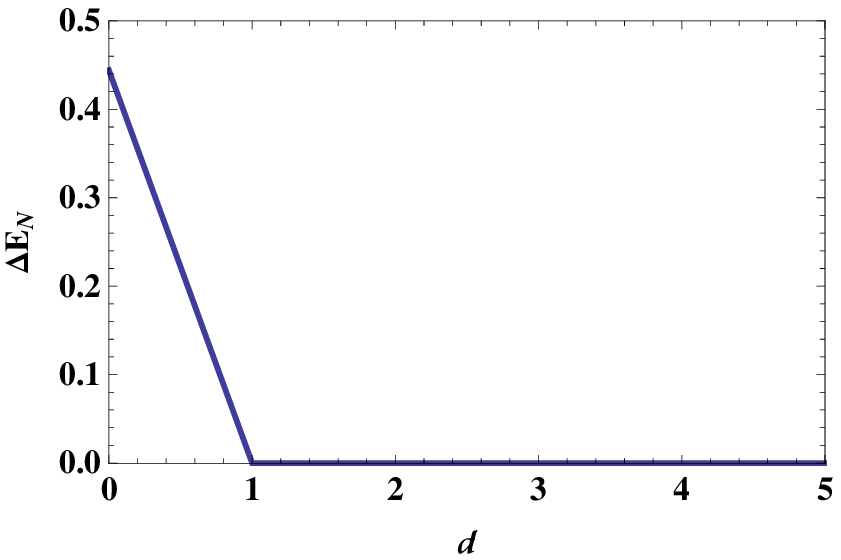}
 \includegraphics[width=0.45\linewidth,clip]{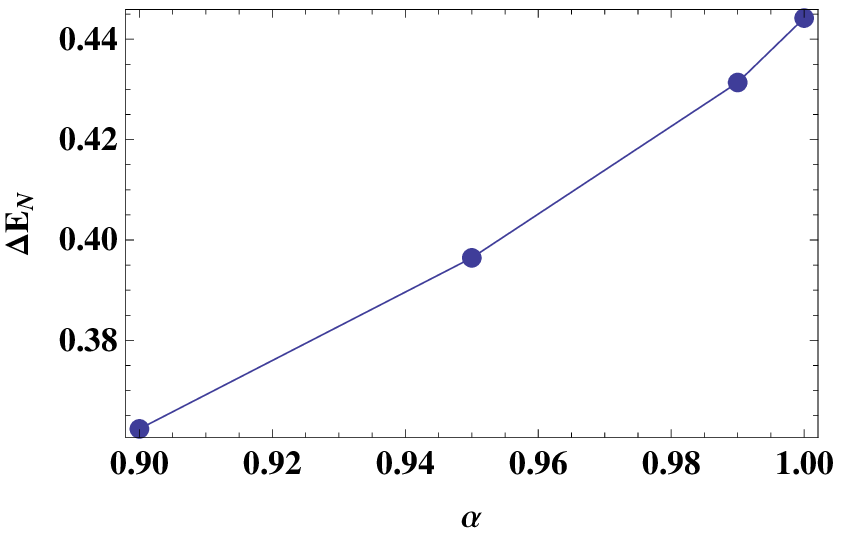}
 \caption{Left panel: $d$ dependence of the decrease of the
   logarithmic negativity due to the measurement for
   $\al_4=1-10^{-7}$. The logarithmic negativity before and after the
   measurement is zero for $d\ge 1$ and this behavior is the same for the
   other values of $\al$. Right panel:  $\al$ dependence of the
   $\Delta E_N$ for $d=0$.}
  \label{fig:neg-set2}
\end{figure}
\noindent
Figure~\ref{fig:neg-set2} shows $\Delta E_N$ as a function of the
separation between $A$ and $B$. The bipartite system composed of $A$
and $B$ is entangled only for $d=0$ before the measurement. After the
measurement, $A$ and $B$ become separable even for $d=0$.  This
behavior is the same for other values of $\al$.  As $\al$ approaches
1, the entanglement consumed via measurement increases. Our numerical
calculation indicates that we have no entanglement breaking for
$d=1,2,3,\cdots$. However, $\Delta E_N=0$ for $d\neq 0$ does not mean
sites $A$ and $B$ do not have the correlation necessary to establish the QET
protocol.
\begin{figure}[H]
  \centering
  \includegraphics[width=0.5\linewidth,clip]{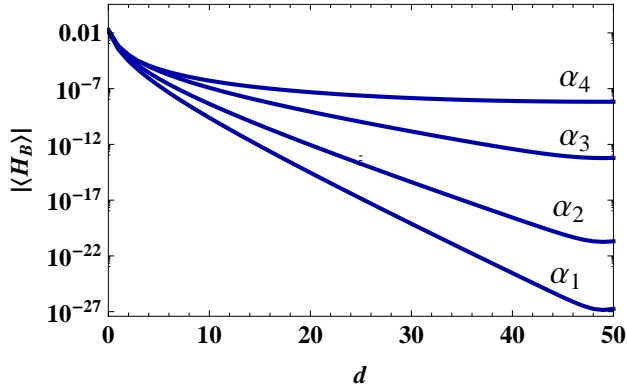}
  \caption{$d$ dependence of the optimized energy of the site $B$.
    For the critical case $\al=\al_4$, this quantity behaves as
    $|\langle H_B\rangle|\sim 2\times 10^{-3}d^{-3.6}$ for large $d$.}
  \label{fig:ene-set2}
\end{figure}
\noindent
Indeed, as Fig.~\ref{fig:ene-set2} shows, we have non-zero (negative)
optimized energy of the site $B$ even for $\Delta E_N=0$ ($d\ge 1$). The
amount of  optimized energy of site $B$ decreases as $d$ increases.
For the critical case $\al_4$, the optimized energy for large $d$
behaves as
\begin{equation}
  \label{eq:scaling1}
  |\langle H_B\rangle|\sim 2\times 10^{-3}d^{-3.6}.
\end{equation}
To understand the behavior of the optimized energy from the view point of
entanglement breaking, we investigated the $d$ dependence of the mutual
information \eqref{eq:mutu} (Fig.~\ref{fig:info-set1}).
\noindent
The mutual information represents the total correlation of the
bipartite system including the quantum and the classical
correlations~\cite{HendersonL:JPAMG34:68996905}. This quantity is
suitable for understanding the behavior of QET for the setting 1
because the decrease of the logarithmic negativity becomes zero for
$d\ge 1$, and it seems that we do not have any quantum correlations
between $A$ and $B$. In this case, the information on the entanglement
of the ground state is encoded as classical correlation between $A$
and $B$, which is established after the POVM measurement of $A$. As
$d$ increases, the mutual information monotonically decreases.  For
the critical case, we have
\begin{equation}
  \label{eq:scaling2}
     \Delta S_M\sim    1.55d^{-0.11}.
\end{equation}
The behavior of the mutual information is consistent with the $d$
dependence of the optimized energy; the larger the amount of the
mutual information, the larger the amount of the optimized
energy. Thus, we can conclude that the amount of energy teleported 
via the protocol of QET is related to the amount of  breaking of
the correlation between the group $A$ and $B$. Here it should be stressed
that the ``classical'' correlation supporting energy teleportation is
induced by ground-state entanglement which is purely quantum. If we
have no entanglement in the ground state, we do not have any
correlation between $A$ and $B$.

For the critical case, we obtained the scaling behavior
\eqref{eq:scaling1} and \eqref{eq:scaling2} numerically. However, we
have not developed any theoretical explanation for the values of these
power law indices. We expect that an analysis based on  conformal
field theory may reveal these values.
\subsection{Setting 2}
\begin{figure}[H]
  \centering
  \includegraphics[width=0.5\linewidth,clip]{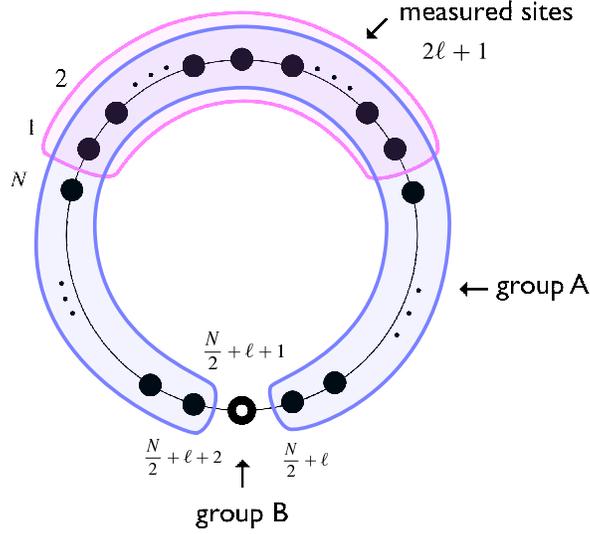}
  \caption{The setting 2 of our numerical calculation. We change the
    number of measured sites in  group $A$.}
  \label{fig:setting2}
\end{figure}

In this setting, the group $A$ consists of $N-1$ sites and the group $B$
is its complement and consists of a single site
(Fig.~\ref{fig:setting2}).
\noindent
We change the number $2\ell+1$ of  measured sites in  group
$A$. As we  explained in Sec.~III, the procedure of the measurement
infuses energy into the system. Hence, as $\ell$ increases, the state of
the bipartite system after the measurement is strongly affected by the
measurement process. In this sense, we can control the strength of the
measurement from ``weak'' to ``strong'' by increasing the number of
measured sites $\ell$ in the group $A$. We expect that the stronger the
measurement becomes, the larger the amount of  entanglement
breaking and this enables us to obtain a larger amount of teleported 
energy of the site $B$ via the protocol of QET.

\begin{figure}[H]
  \centering
  \includegraphics[width=0.47\linewidth,clip]{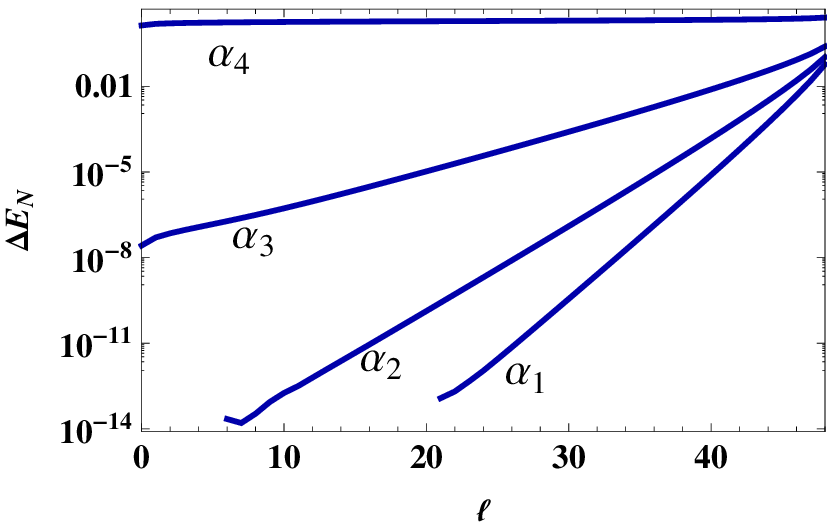}
 \includegraphics[width=0.47\linewidth,clip]{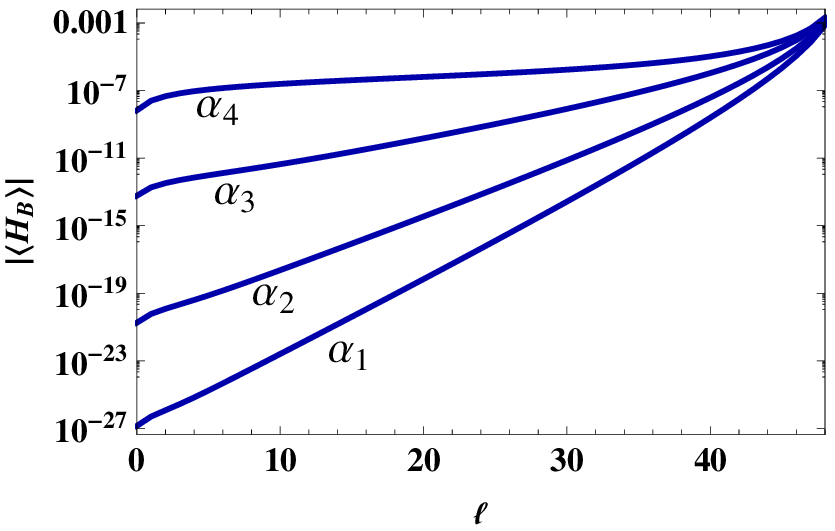} 
  \caption{Left panel: decrease of the logarithmic
    negativity due to the  measurement. Right panel: 
    optimized energy of site $B$.}
  \label{fig:neg}
\end{figure}
\begin{figure}[H]
  \centering
  \includegraphics[width=0.6\linewidth,clip]{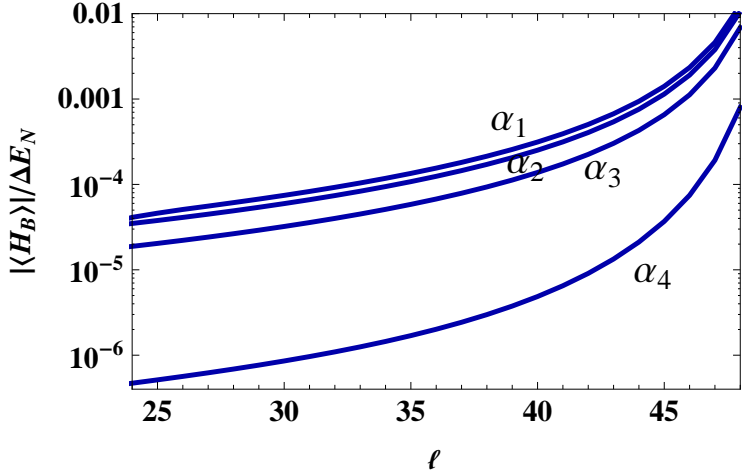}
  \caption{Ratio of the optimized energy of site $B$ and the amount
    of entanglement breaking.}
  \label{fig:ratio1}
\end{figure}

Figure~\ref{fig:neg} shows the decrease of the logarithmic negativity due to
the measurement, and the optimized energy obtained for the site $B$ via
QET for the system size $N=100$. We have also investigated the system
sizes $2\le N\le 100$ and confirmed that the behavior of these
quantities is same. Hence, we present the result with $N=100$.  As
the measurement becomes ``strong'' ($\ell$ increases), the amount of
 entanglement breaking and the resulting optimized energy of site $B$
increases. In this setting, as the total bipartite system is pure
state, the system becomes separable after the measurement when the all
sites in $A$ are measured and the entanglement breaking due to the
measurement is maximal.
This behavior is consistent with our naive expectation that the ground-
state entanglement of the harmonic chain is a resource for
QET. Figure~\ref{fig:ratio1} is the $\ell$ dependence of the ratio of the
optimized energy of site $B$ and the entanglement breaking. 
\noindent
As we can observe, this ratio is monotonically increasing with respect
to the size of the measured site $\ell$ and bounded from above. The
maximum value of the ratio is achieved at $\ell=N/2-2$, which is the maximal number of
the measuring sites in  group $A$. Thus, the maximal amount of 
teleported energy is bounded by the amount of entanglement breaking
and the following inequality holds:
\begin{equation}
  |\langle H_B\rangle|<\beta \Delta E_N,\quad \beta <1,
\end{equation}
where $\beta$ is a constant whose value depends on the system size
$N$ and the parameter $\al$. To find the relation between the
amount of  entanglement breaking and the optimized energy, we
investigated the $N$ dependence of $\Delta E_N$ and $\langle H_B\rangle$
for a specific value of $\ell=N/2-2$; this value of $\ell$ corresponds
to the maximal possible number of  measuring sites in  group $A$
for a given $N$. The entanglement breaking due to the measurement is
expected to become maximal (the strongest measurement case).

\begin{figure}[H]
  \centering
  \includegraphics[width=0.48\linewidth,clip]{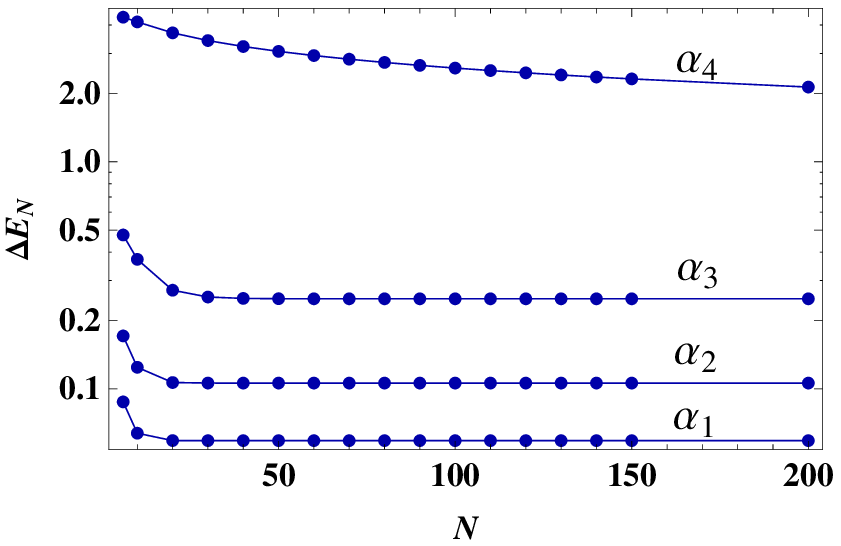}
  \includegraphics[width=0.5\linewidth,clip]{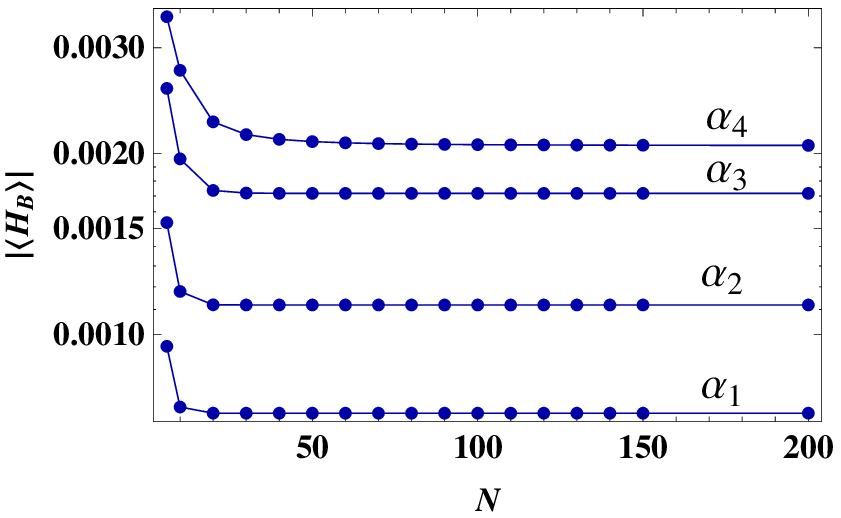}
  \caption{$N$ dependence of $\Delta E_N$ and $|\langle
    H_B\rangle|$ for  $\ell=N/2-2$.}
  \label{fig:depN}
\end{figure}

\begin{figure}[H]
  \centering
  \includegraphics[width=0.6\linewidth,clip]{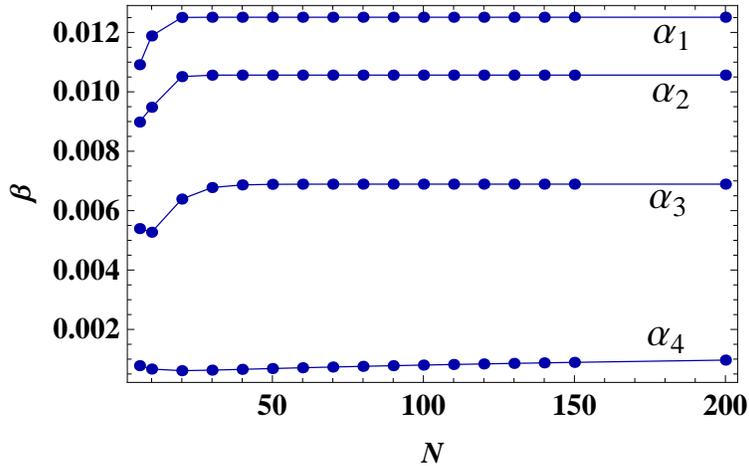}
  \caption{The function $\beta(N)$.}
  \label{fig:beta}
\end{figure}

As the system size $N$ increases, the amount of  entanglement
breaking and the teleported energy both decrease (Fig.~\ref{fig:depN}).
\noindent
For the non-critical case $\al=\al_1,\al_2,\al_3$, these quantities
approach constant values for large $N$.     For the critical case
$\al=\al_4$, we have the following scaling behavior
\begin{equation}
 \Delta E_N\sim 8N^{-0.32},\qquad |\langle H_B\rangle|\sim 0.12 N^{-2.1}+0.0020613.
\end{equation}
By taking the ratio of these quantities, we found that the following
inequality holds:
\begin{equation}
  |\langle H_B\rangle|<\beta(N)\Delta E_N.
  \label{eq:ineq}
\end{equation}
\noindent
The behavior of the function $\beta(N)$ is shown in Fig.~\ref{fig:beta}. 
\noindent
For $\al=\al_1,\al_2,\al_3$, $\beta$ asymptotically approaches
constant values. For the critical case $\al=\al_4$, $\beta$ behaves
as
\begin{equation}
  \beta(N)\sim 0.00026N^{0.32}.
\end{equation}
The inequality \eqref{eq:ineq} implies that a large amount of
teleported energy requires a large amount of consumption of the
ground- state entanglement between the groups $A$ and $B$. In other
words, the possible amount of  teleported energy is bounded by the
amount of  entanglement breaking. This relation was previously
confirmed for the minimal QET model by  one of
authors~\cite{HottaM:PLA374:2010}. We have confirmed that a similar
relation holds for the QET model with a harmonic chain.

It is noted that our bound \eqref{eq:ineq} with $N=2$ does not
quantitatively coincide with the bound of \cite{HottaM:PLA374:2010}
because we treat not a qubit chain as in  \cite{HottaM:PLA374:2010}, but a
harmonic chain. However, there also exists a qualitative difference
between our result with $N=2$ and the result in
\cite{HottaM:PLA374:2010}. Due to noncommutativity between the interaction
Hamiltonian and measurement operators, our measurement of $A$ disturbs
energy of $B$ directly when $N=2$. Thus the energy gain from $B$ is not
purely interpreted as QET output. (However, when $N\ge 4$, the
measurement of $A$ does not change energy of $B$; thus, the energy gain
from $B$ relies only on QET.) In \cite{HottaM:PLA374:2010}, the
measurement of one qubit is a non-demolition measurement for the other
qubit. Thus, even in the case of two qubits, the energy gain of
\cite{HottaM:PLA374:2010} is transported by QET.

\section{Summary}

We numerically investigated the protocol of QET for a harmonic chain
model. For a bipartite system in the harmonic chain, we applied the
QET protocol defined via a POVM measurement and  LOCC. The
resource for QET in the harmonic chain is the ground-state
entanglement between spatially separated two groups. We can extract
energy from the system by breaking the entanglement via the POVM
measurement and following LOCC.

We confirmed that the amount of extractable energy via the protocol of
QET is bounded from above by the amount of the entanglement consumed
due to the POVM measurement. This implies that a large amount of
teleported energy needs a large amount of consumption of the
ground-state entanglement between the two groups in the bipartite
system.  We also considered the situation that the logarithmic
negativity between two groups is zero. In this case, the two groups
are separable and there is no quantum correlation between them. Even
 in such a case, we showed that we can extract energy via the protocol of
QED and the amount of energy is correlated with the amount of 
breaking of the mutual information, which quantifies the total
correlation between the two groups. It should be emphasized that the
above ``classical'' correlation supporting QET is originally induced
from a purely quantum correlation, that is, the ground-state
entanglement. If we have no entanglement in the ground state, such
``classical'' bipartite correlation between $A$ and $B$ does not take
place at all. As pointed out above, QET only needs ``classical''
correlation of $A$ and $B$. This result suggests that QET processes are
tolerant of decoherence which destroys the quantum correlation of $A$ and $B$,
unlike the  usual quantum teleportation.

\begin{acknowledgments}
This research has been partially  supported  by the Global COE Program
of MEXT, Japan, and the Ministry of Education, Science, Sports and
Culture, Japan, under Grant No. 22540406.
\end{acknowledgments}



\end{document}